\documentclass[11pt]{article}
\pdfoutput=1
\usepackage{jheppub}

\usepackage{amsmath}

\usepackage{float}

\newcommand\be{\begin{equation}}
\newcommand\ee{\end{equation}}
\newcommand{\bea}{\begin{eqnarray}}
\newcommand{\eea}{\end{eqnarray}}

\newcommand\mS{{\mathbb S}}
\newcommand\mR{{\mathbb R}}

\newcommand\mH{{\mathbb H}}

\usepackage{color}
\newcommand\numberthis{\addtocounter{equation}{1}\tag{\theequation}}
\numberwithin{equation}{section}

\def\<{\langle}
\renewcommand\>{\rangle}

\newcommand{\Tr}{{\rm Tr}}

\title{Hyperbolic Lattice for Scalar Field Theory in AdS$_3$}

\author[a,b]{Richard C. Brower,}
\author[a]{Cameron V. Cogburn,}
\author[a]{and Evan Owen}
\affiliation[a]{Department of Physics, Boston University,\\
  590 Commonwealth Avenue, Boston, MA 02215, USA}

\affiliation[b]{Center for Computational Science, Boston University,\\
  3 Cummington Mall, Boston, MA 02215, USA}

\emailAdd{cogburn@bu.edu}
\emailAdd{brower@bu.edu}
\emailAdd{ekowen@bu.edu}

\abstract{We construct a tessellation of AdS$_3$, by extending the equilateral  triangulation  of AdS$_2$ on the Poincar\'{e} disk based on the $(2,3,7)$ triangle group,
suitable for studying strongly coupled phenomena and the AdS/CFT correspondence. 
A Hamiltonian form conducive to the study of dynamics and quantum computation is presented. 
We show agreement between lattice calculations and analytic results for the free scalar theory and find evidence of a second
order critical transition for $\phi^4$  theory using Monte Carlo simulations. 
Applications of this AdS Hamiltonian formulation to real time
evolution and quantum computing are discussed. 
 }

\begin{document}
\maketitle
\parskip=12pt

\section{Introduction} \label{sec:Intro}

The study of strongly coupled Quantum Field Theories (QFTs) is
difficult, even before putting them in anti-de Sitter (AdS) space. Yet there are compelling reasons to understand the full non-perturbative
consequences~\cite{Callan:1989em} of doing so.  The general AdS/CFT duality maps any ultraviolet (UV)
renormalizable field theory in bulk AdS$_{d+1}$
to a $d$-dimensional Conformal Field Theory (CFT$_d$) on the boundary.
Despite the spectacular results of the conformal bootstrap program  in bounding
CFTs with a conserved stress tensor (e.g., \cite{Rattazzi:2008pe, 
El-Showk:2012cjh}), there is still a huge landscape of 
non-perturbative  field theories in AdS yet to be explored with potential
applications to particle and condensed matter physics. 
Moreover, in three dimensions AdS$_3$
gives the interesting and special case of a two-dimensional boundary CFT$_2$.
It is also the minimum dimensionality needed to study
the map between non-trivial pure weak gravity and CFT with a conserved 
stress tensor. 

The best developed numerical framework for solving  strongly coupled QFTs at the moment is lattice field theory.
This method has benefited from decades of development of efficient algorithms and high performance architectures  to produce
extremely precise predictions for 
physical systems such as Quantum Chromodynamics \cite{ExascaleWP}. To best utilize this framework,
lattice QCD and similar problems are posed  as a path integral in flat Euclidean space, which benefits from a regular lattice with a uniform UV cut-off  and
 a positive definite  measure allowing for  the efficient use of parallelized Monte Carlo algorithms.
 
 Adapting  lattice field theory to AdS space presents two new problems: (i)  lattices  must conform 
to a curved manifold and (ii) the finite lattice volume must have a boundary that maps to a CFT at infinite distances.
The first problem  (i) has  largely been addressed in  the Quantum Finite Element  (QFE) program
by introducing  a simplicial lattice
complex  weighted by the discrete exterior calculus \cite{Brower:2016vsl,Brower:2018szu,Brower:2020jqj,brower2021PoS}.  QFE is
proving to give accurate results for the
$\phi^4$ Ising CFT in 2D on the Riemann sphere $\mS^2$ and in 3D for the radially quantized cylinder $\mR \times \mS^2$.

The extension of QFE to a hyperbolic manifold was presented in \cite{Brower:2019kyh} for AdS$_2$.
In addition,  for  AdS$_2$ the second problem (ii) of  convergence  as a function of  
the UV cut-off to the boundary CFT was shown to be feasible at finite volumes.  
 In this work we extend this investigation by choosing a foliation for global
Euclidean AdS$_{d+1}$ that defines the  Hamiltonian (dilatation) operator dual to the boundary CFT 
in the radially quantized formulation on $\mR \times \mS^d$. In 3D this particular AdS$_3$ 
geometry allows for the re-use of the basic lattice scaffolding  of the 
Poincar\'{e} disk $\mH^2$ by tessellating each 2D slice via the triangle group at fixed time.  This lattice field theory
approach to the non-perturbative study of QFTs in AdS space is complementary to the S-matrix bootstrap approach \cite{Paulos:2016fap, Paulos:2016but} as well as 
increasingly powerful Hamiltonian truncation methods \cite{Rychkov:2015vap, Hogervorst:2021spa,  
  Anand:2020qnp}.  The use of the Hamiltonian formulation of
lattice field theory opens up potential  applications to Minkowski space complementary to
the light-cone truncation method \cite{Katz:2016hxp, Anand:2020gnn} and generally to quantum computing algorithms.

This article begins in Sec.~\ref{sec:Setup} with a general discussion of
the AdS manifold and our  lattice construction of
AdS$_3$.  Section \ref{app:HamForm} details the Hamiltonian and
Lagrangian formulation of $\phi^4$ theory on the lattice.  Section
\ref{sec:FreeTheory} focuses on the free theory, where we compute
various propagators directly to compare the lattice to the continuum
and as a check of our Monte Carlo methods. Section \ref{sec:MC} then uses 
these methods to demonstrate evidence for the existence of a  second order critical point in
$\phi^4$ theory in AdS$_3$.  We finish in Sec.~\ref{sec:discussion} with a discussion of future directions for high
precision Monte Carlo simulations for both  $\phi^4$ and Ising spins in
AdS$_3$ to probe the AdS/CFT correspondence,
as well as using the Hamiltonian form as a prototype for quantum 
computing algorithms.  

\section{Anti-de Sitter Space} \label{sec:Setup}

We begin with a general discussion about AdS space and its various foliations before proceeding to its latticization. 
Euclidean AdS$_{d+1}$ with curvature radius $\ell$ is a space of constant negative curvature 
defined as the hyperboloid,
\be
-  X^2_0 +  \vec X \cdot \vec X    = - X_0  X_0 +  \sum_{i=1}^{d+1} X_{i} X_{i}   = -\ell^2 \; ,
\quad \quad X_0 > 0 \; ,
\ee
embedded in $\mathbb{R}^{1,d+1}$ and possessing
the isometries
of the Euclidean conformal group $SO(1,d+1)$. 
Between any two points $X$, $X'$ on the manifold, there exists
a unique geodesic given by
\be
\ell^2\cosh(\sigma(X,X'))  =   X_{0} X'_{0} - \vec X \cdot \vec X'  \ge 0 \; ,
\ee 
which, when projected onto the hyperbolic surface 
 is a positive spacelike distance
as seen by the alternative expression
$\ell^2\sinh^2(\sigma/2) = (\vec X -\vec X')^2 - (X_0 -X'_0)^2 $.

\begin{figure} 
\begin{center}
\includegraphics[width=\textwidth, angle=0, scale=1.0]{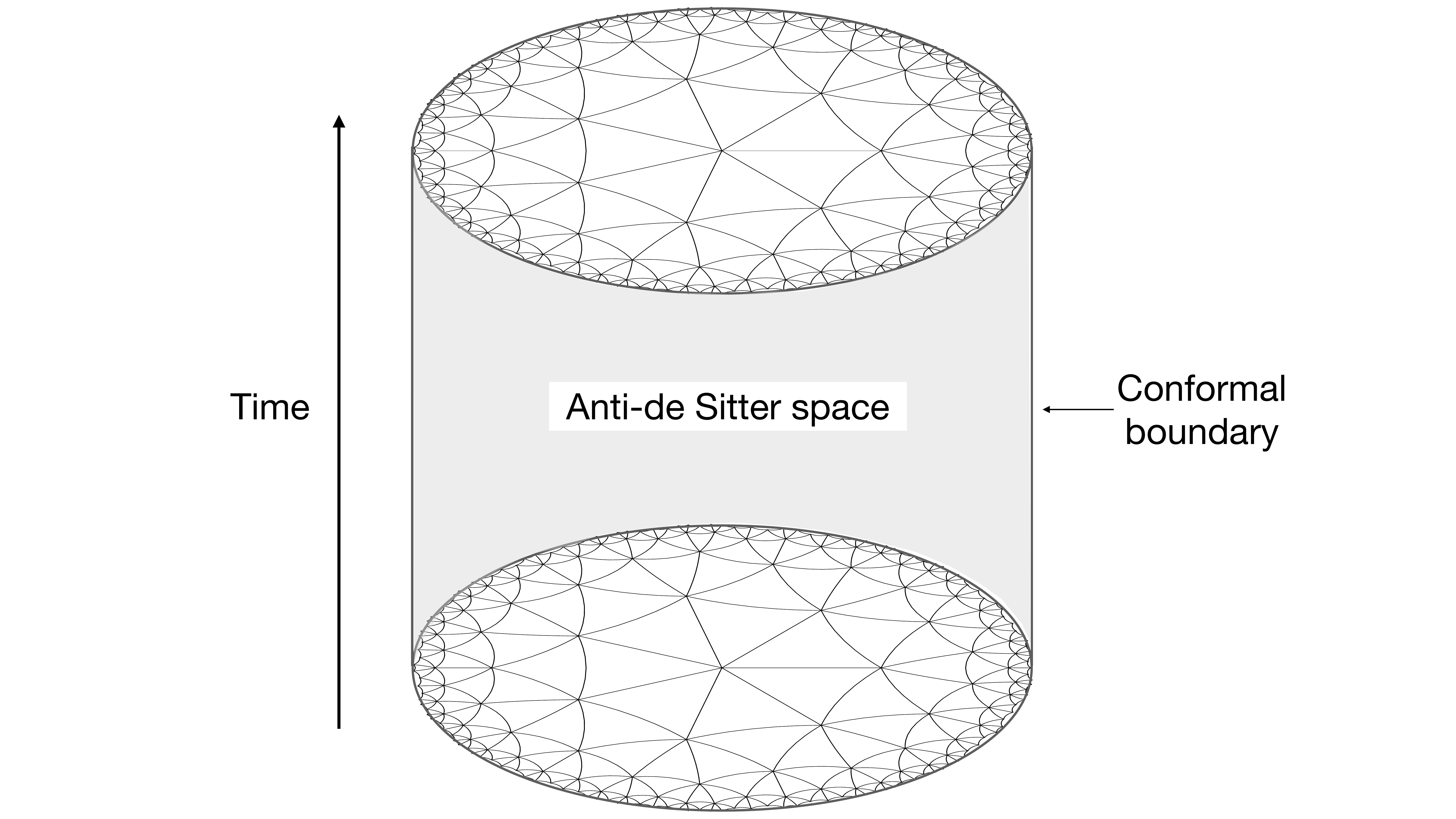}
\caption{AdS$_3$ spacetime with our choice of coordinate system looks like a solid cylinder. At fixed time the space is 
the hyperbolic disk, which can be tessellated using equilateral hyperbolic triangles. 
Here, $(2,3,7)$ triangles are used.}
\label{fig:AdSCylinder}
\end{center}
\end{figure}

Before constructing a lattice it is useful to think about the choice of coordinates 
taken on the hyperbolic surface \cite{Bengtsson98}.
Three conventional ones are the Upper Half-Plane (UHP), the Poincar\'{e} ball, and
the AdS cylinder (see Fig.~\ref{fig:AdSCylinder}). The boundary CFTs for these three coordinate systems are on different
manifolds: Euclidean $\mR^d$, the sphere $\mS^d$, and the
cylinder $\mR \times \mS^{d-1}$, respectively. For each choice, the hyperbolic manifold remains
 unchanged while  
the boundary CFT maps to
different manifolds related by Weyl factors. 
This well known fact is emphasized by Witten \cite{Witten:1998qj} but is sometimes
obscured by referring to both the Weyl equivalences and isometries of AdS space as ``conformal".

Euclidean AdS$_{d+1}$
has the topology of a cylinder $\mR \times \mH^{d}$ 
with its metric being the sum of two terms,
\be \label{eq:AdSDmetricBD}
ds^2 = g_{00}dt^2 + ds^2_{\mathbb H^d}  \; ,
\ee
which separates Euclidean time $t \in (-\infty, \infty)$ from 
the spatial metric on $\mathbb H^d$. 
Time translation is generated by the dilatation operator 
$D = -\partial_{t}$ (or AdS Hamiltonian) with unitary evolution
in Minkowski space corresponding to the replacement $t \rightarrow -it$. 
The temporal metric component $g_{00}$ is a function of a radial coordinate
on $\mH^d$, but causal propagation in the
bulk is consistent with causality in the boundary  CFT~\cite{berenstein:2021}.

A particularly nice foliation for AdS
is given by global coordinates, 
\be \label{eq:GlobalCoordsGen}
ds^2 =  \pm \ell^2 \cosh^2 \rho \, dt^2 + \ell^2 (d\rho^2 + \sinh^2 \rho \, d \Omega_{d-1}^2) \; ,
\ee
where  $\rho \in [0,\infty]$ is the geodesic from the origin of $\mH^d$ at
fixed time with   $g_{00}(\rho) = \ell^2 \cosh(\rho)$ and $d\Omega^2_{d-1}$ is the line element of the unit sphere $\mS^{d-1}$.\footnote{In general  $d\Omega^2_{d-1}$  is
determined by the recursion relation  $d\Omega^2_n = d\theta^2_n+ \sin^2(\theta_n)d\Omega^2_{n-1}$.} 
The minus sign is for Minkowski AdS whereas the plus sign gives 
Euclidean AdS.  
For AdS$_3$ this is then
\be \label{eq:globalAdS3metricHYP}
ds^2 = \ell^2 (\cosh^2 \rho \, dt^2 + d\rho^2 + \sinh^2 \rho \, d \theta^2) \; ,
\ee
with $d\Omega^2_1 = d\theta^2$ on $\mS^1$.

In global coordinates the conformal 
boundary is at $\rho = \infty$. 
By compactifying the radial coordinate through $r = \tanh(\rho/2)$ we obtain the Poincar\'{e} disk coordinates,
\be
ds^2 = \frac{\ell^2}{(1-r^2)^2}  \left( (1+r^2)^2 dt^2 + 4(dr^2 + r^2 d\theta^2) \right) \; ,
\ee
which include the conformal boundary at $r=1$. 
This form of the line element makes particularly clear 
the cylindrical topology, i.e., the
time translation and $SO(2)$ symmetry.
In practice, a UV cut-off $\epsilon = 1 -  r_{\text{max}}  = \mathcal{O}(e^{ -\rho_{\text{max}}})$ is introduced for the boundary 
CFT at $r=1$. A picture of AdS$_3$ spacetime is shown in Fig.~\ref{fig:AdSCylinder}.

We note that for AdS$_2$ with  $d=1$, the cylindrical form of the metric is reduced to  an infinite strip 
via the change of variables $\cosh\rho = 1/\cos(\sigma)$. 
The metric (\ref{eq:GlobalCoordsGen}) is then given by
\be \label{eq:strip}
ds^2 = \ell^2 (\cosh^2 \rho \, dt^2 + d\rho^2 ) =  \frac{\ell^2}{\cos^2 \sigma}  [dt^2 + d\sigma^2] \; ,
\ee
with $\sigma \in [-\pi/2, \pi/2]$ and the 1D conformal quantum mechanics exists on the boundary at
$\sigma = \pm \pi/2$. This form is 
convenient for the Hamiltonian truncation methods
presented in \cite{Hogervorst:2021spa}.

\begin{figure} 
\includegraphics[width=0.45\textwidth, angle=0, scale=0.99]{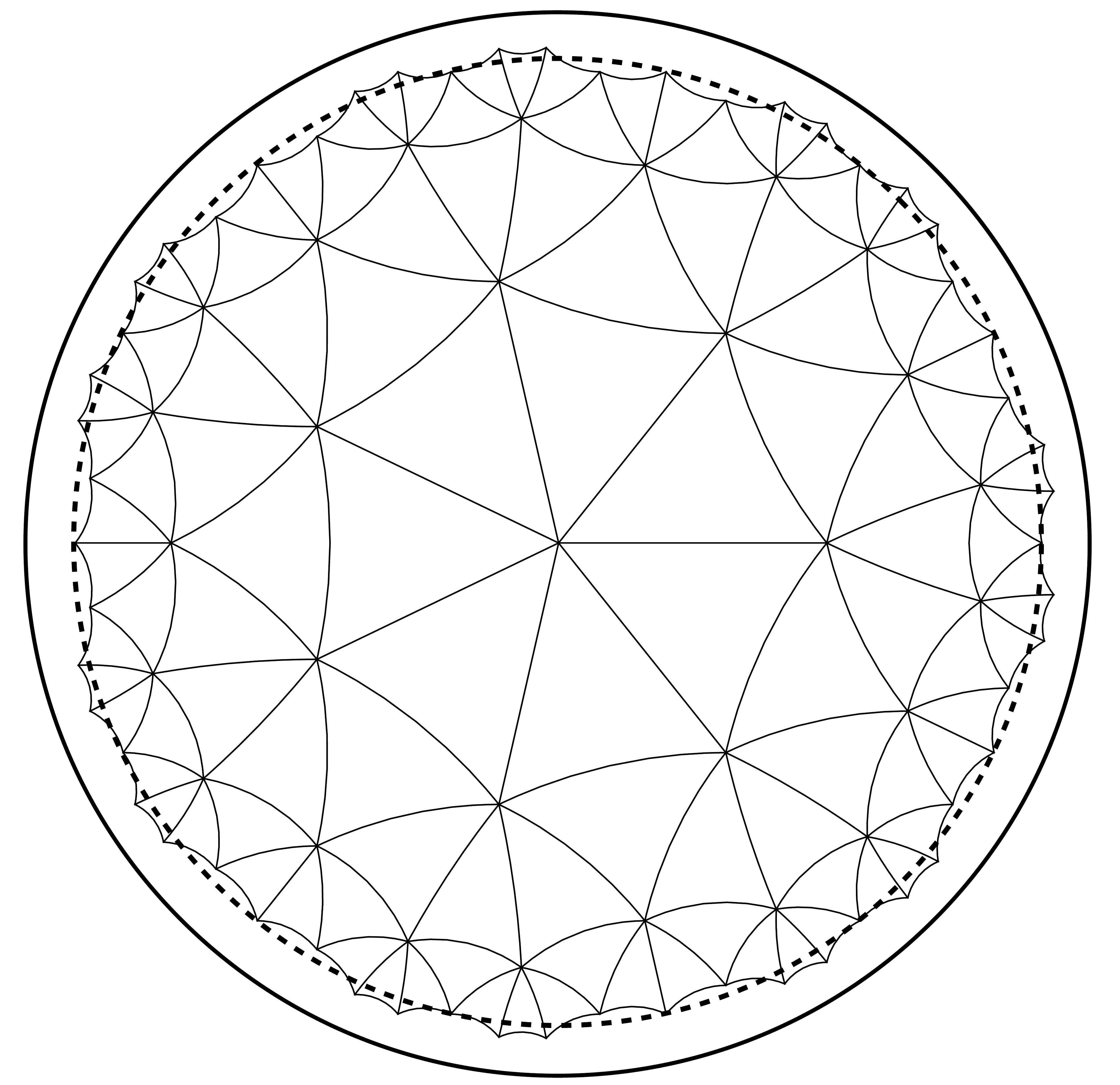}
\hspace{1cm}
\includegraphics[trim=0 -5.5cm 0 0cm, width=0.55\textwidth, angle=0, scale=0.99]{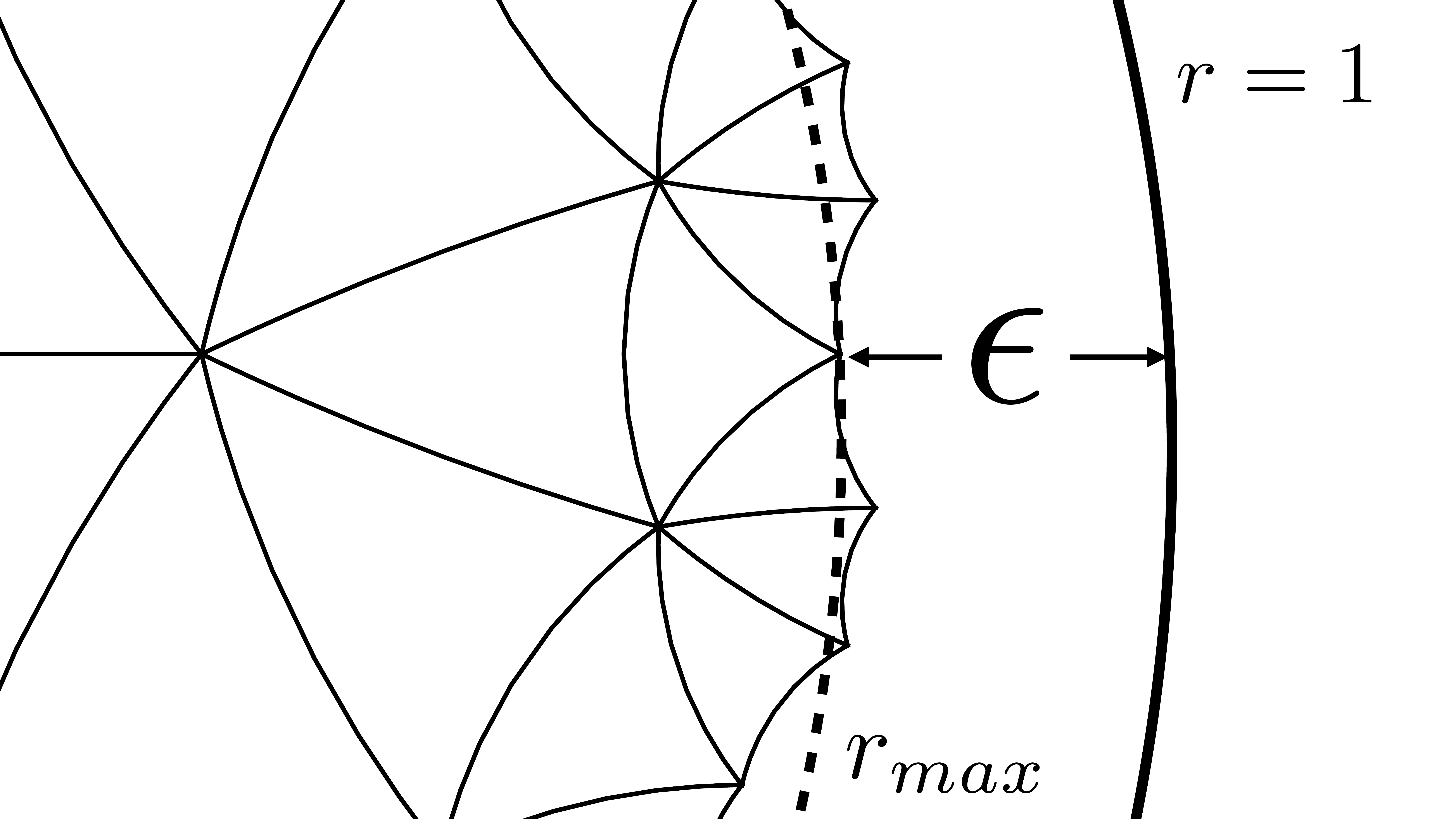}
\caption{The full hyperbolic disk $\mH^2$ and a zoomed 
in view of the edge,
tessellated with three layers $(L=3)$ of $(2,3,7)$ triangles. 
The dashed circle near the lattice edge is the effective lattice boundary $r_{\text{max}}$ and 
 the larger, solid circle the conformal boundary
 at $r=1$. The UV lattice cut-off is $\epsilon = 1 - r_{\text{max}}$.
}
\label{fig:LatticeEdge}
\end{figure}

\subsection{Choosing the spatial lattice} \label{sec:spatiallatice}

A lattice field theory calculation replaces the
continuum manifold with a finite lattice spacing $a$ (UV cut-off)
and a finite  volume  $V$ (IR cut-off). 
For example,
for the $d$-dimensional flat space cubic lattice with toroidal boundary conditions
with $V = \mathcal{O}(L_1L_2  \cdots  L_d)$ and $\mathcal{O}(L_\mu/a)$ lattice sites on each axis, 
the lowest mass  $m \sim 1/\xi$ (or gap) must obey $a \ll \xi \ll L_\mu$ in the numerical
extrapolation to the continuum, $a \rightarrow 0,~L_\mu \rightarrow \infty$. 
In principle, to obtain 
the correct boundary CFT  
these limits precede the conformal limit $\xi \rightarrow \infty$.
Since AdS space contains an intrinsic radius of curvature, $\ell$,
we can only access the critical point with the proviso that $a \ll \ell  \ll \rho_{\text{max}}$. 

Given the orthogonality in the metric (\ref{eq:AdSDmetricBD}),
it is natural to foliate Euclidean AdS into fixed time slices transverse to the spatial
 $\mH^d$ metric
and then subsequently introduce a spatial lattice for each time slice. 
For AdS$_3$ the spatial tessellation is identical to the
lattice realization for $\mathbb H^2$ detailed in \cite{Brower:2019kyh}.
The hyperbolic disk can then be tessellated using  equilateral triangles
of the $(2,3,q)$ triangle group for $q > 6$. Throughout this work we use $q=7$ because it gives the smallest equilateral hyperbolic triangle edge length (relative to $\ell$), which minimizes the curvature defects at the vertices. A complementary way to latticize AdS$_3$ using a regular tessellation of $\mH^3$ is done in \cite{Asaduzzaman_2020}.

This construction, not unlike the tessellation of the
flat plane with equilateral triangles $(q = 6)$, gives an infinite lattice with a discrete subgroup of the AdS$_2$ isometries: invariance under translations along
the edges and q-fold rotations about each vertex. 
On a hyperbolic lattice, these symmetries are then broken by
finite volume effects when introducing an IR
cut-off $\rho_x \le \rho_{\text{max}}$ with
an arbitrary center at $\rho = 0$.  Using the triangle group,
the lattice spacing is now fixed relative to the curvature.
For example, for $(2,3,q)$  the deficit angle fixes the equilateral triangle area to 
$A_{\Delta} = (\pi - 6 \pi/q)\ell^2$ and the lengths of the triangles to 
$\cosh(a/2\ell) = \left(2 \sin(\pi/q) \right)^{-1}$.
For $q=7$ this gives the minimum values
$A_{\Delta} = 0.448799 \ell^2$ and $a = 1.090550\ell$. 
In principle, using the finite element method (FEM) \cite{StrangFix200805} each  triangle can be subdivided into $n^2$  flat equilateral triangles with edges $a/n$ subsequently
projected onto the hyperbolic surface using the
same QFE procedure \cite{Brower:2016vsl,Brower:2018szu,brower2021PoS} for the 2D de Sitter manifold $\mS^2$.

In practice, the finite volume is tessellated layer-by-layer from the origin at $\rho = 0$. 
There is an exponential growth in the number of points on the lattice boundary 
as a function of the number of spatial ``layers" $L$ of the lattice, $n_{\text{bdry}} \sim e^{\rho} \sim e^{L}$, 
as expected from holography.
The effective UV lattice boundary cut-off is given by $\epsilon \simeq 2 e^{ - 0.97 \ell L}$, where  
we assume an average $r_{\text{max}}$ for the lattice boundary as opposed to the actual jagged, position-dependent 
lattice boundary generated by this construction (see Fig.~\ref{fig:LatticeEdge}). 
We note that if we were to introduce FEM refinement this would not change the UV cut-off on the boundary and incurs
only a polynomial growth in the number of sites.

The boundary field $\tilde{\phi}$ with scaling dimension $\Delta$ is defined as
\be
\tilde{\phi}(x) = \epsilon^{-\Delta} \phi(x,\sigma) \; .
\label{eq:CFTfield}
\ee
Here $x$ are boundary coordinates and  
$\sigma$ is the  geodesic distance from  the center. We then impose Dirichlet boundary conditions. This suppresses the leading 
term $\tilde{\phi}_0$ of the field
\be
\phi(x,\sigma )  = e^{ -\Delta_+ \sigma} [\tilde{\phi}_1(x) + \mathcal{O}(e^{ - \sigma})] +  e^{ -\Delta_- \sigma} [\tilde{\phi}_0(x) + \mathcal{O}(e^{ - \sigma})] \; ,
\ee
leaving $\tilde{\phi}_1$ as the dynamical fluctuations. For the free field,
 $\Delta_{\pm} = (d/2) \pm \sqrt{(d/2)^2 + m^2}$, and $\tilde{\phi}_0$ and $\tilde{\phi}_1$ are boundary sources.
The scaled field of the boundary
 CFT  is then $\tilde{\phi}_1(x) = \lim_{\sigma \rightarrow \infty} e^{ \Delta_+ \sigma}\phi(x,\sigma )$ 
 or $\tilde{\phi}_1(x) = \lim_{\epsilon \rightarrow 0} \epsilon^{-\Delta_+}\phi(x,\epsilon)$.

The scaling relationship is thus accurate to $\mathcal{O}(\epsilon)$. However, as seen in Table 
\ref{table:DiskInfo}, moderate volumes from just a few layers $L$ give quite 
small values for the UV cut-off $\epsilon(L)$, from which we can then extrapolate 
to zero to identify boundary phenomena. Specifically for $q = 7$,
as L increases the number of nodes on the disk
grows exponentially as $N(L) \simeq 5.086 e^{0.96 L}$ relative to the number of nodes on the outer edge $E(L) \simeq 3.13 e^{0.96 L}$ with the ratio
approaching the inverse of the golden ratio, $E(L)/N(L) \rightarrow (1 - \sqrt{5})/2  = 0.618034\dots$ as the number of layers increases.

\begin{table}[h!]
\centering
\begin{tabular}{|c|c|c|c|c|c|c|c|c|c|c| }
\hline
Layers $L$  & 0 & 1 & 2 &3 &4 & 5 & 6 & 7 & 8     \\  
\hline
Disk Nodes $N(L)$    & 1 & 8 & 29 & 85 & 232 & 617& 1675 & 4264 & 11173   \\
\hline 
 Edge  Nodes $E(L)$& 1 & 7 & 21 & 56  &147 &385 & 1008 & 2639 & 6909  \\ 
\hline 
UV cut-off $\epsilon$   & 1 & 0.50  & 0.23 & 0.097 & 0.038 &  0.015 &  0.0057  & 0.0022 & $8.3 \times 10^{-4}$  \\ 
\hline 
\end{tabular}
\caption{The total number of nodes on the disk $N(L)$, 
the number of edge nodes $E(L)$ on the outside layer, and the associated UV cut-off $\epsilon = 1 - r_{\text{max}} $ 
as a function of the number of the number of layers $L$ with $q=7$.}
\label{table:DiskInfo}
\end{table}

\section{AdS$_3$ Hamiltonian for $\phi^4$ theory } \label{app:HamForm}

To study $\phi^4$ theory on 
a simplicial triangulation of $\mH^2$ with continuous time $t$, we
begin with the continuum action
\be \label{eq:phi4action}
S = \int dt \int d^2x \sqrt{g} \left( \frac{1}{2} g^{\mu \nu} \partial_{\mu} \phi \partial_{\nu} \phi + \frac{1}{2} m^2 \phi^2 + \lambda \phi^4 \right)  \; .
\ee
The resulting spatially discretized action is 
\begin{multline*} \label{eq:discreteAction}
S = \int dt  \sum_x \Bigg[ \frac{1}{2} \sum_{y \in \<x,y\>}  \dfrac{\cosh \rho_x}{2} K_{xy}
(\phi_{x} - \phi_{y})^2 \\
+ \sqrt{g_x} \cosh \rho_x  \left( \frac{1}{2 \cosh^2 \rho_x} (\partial_t
\phi_x)^2 +  
\frac{1}{2} m^2 \phi^2_{x} + \lambda \phi^4_{x} \right) \Bigg] \; , \numberthis
\end{multline*}
where the notation $y \in \<x,y\>$ indicates a sum over all nearest neighbors of site $x$ in the AdS$_2$ graph,
and we have inserted the discretized metric coefficients $\sqrt{g} = \cosh \rho_x \sqrt{g_x} $ and $  g^{00}_x =
1/\cosh^2 \rho_x$.
The coefficients $\sqrt{g_x}$ and $K_{xy}$ can be determined using the FEM.
This method sets the weight of each site $\sqrt{g_x}$ to the volume of the dual site, and the kinetic weight of each link $K_{xy}$ to the ratio of the dual link length (the Hodge star of the link) to the length of the link itself.

At present we do not introduce further QFE refinement. Therefore the  weights $\sqrt{g_x}$ and $K_{xy}$ are constant,
\be \label{eq:2dweights}
\sqrt{g_x} = \frac{q}{3} A_{\Delta} \, , \quad \quad K_{xy} = \frac{4 A_{\Delta}}{3 a^2} \, ,
\ee
with the lattice space $a$ given by
\be
\cosh(a/2\ell) = \frac{\cos(\pi/3)}{\sin(\pi/q)} = \frac{1}{2 \sin(\pi/q)} \; .
\ee
As mentioned in Sec.~\ref{sec:Setup}, the curvature defects on the lattice are minimized
for $q = 7$  with the minimal lattice spacing $a = 1.090550\ell $.
Although this might suggest that refinement is necessary to get sensible results, it
was shown in \cite{Brower:2019kyh} that excellent 
long-distance
propagators 
 can be obtained without refinement.  By virtue of the
IR/UV map in the AdS/CFT correspondence, this even
suggests the possibility that this discrete lattice might give exact continuum
CFTs on the boundary in the infinite volume limit. 

On  the infinite lattice there exists a Hamiltonian form equivalent to
the lattice Lagrangian (\ref{eq:discreteAction}),
\begin{multline*} \label{eq:discreteHam}
\hat H = \sum_x \left[ \frac{1}{2} \sum_{y \in \<x,y\>}  \dfrac{\cosh \rho_x}{2} K_{xy}
(\hat \phi_{x} - \hat \phi_{y})^2 +
\sqrt{g_x} \cosh \rho_x  \left( \frac{1}{2} \hat \pi^2_x +
\frac{1}{2} m^2 \hat \phi^2_{x} + \lambda \hat \phi^4_{x} \right) \right] \; , \numberthis
\end{multline*}
which  avoids the subtleties associated with 
discretizing time while keeping a regular tessellation of the disk. 
The operators obey the canonical commutation relation
\be \label{eq:MomComm}
[\hat \phi(x),\hat \pi(x') ] = i   \frac{\delta^2(x-x')}{\sqrt{g(x)}}
\quad \rightarrow \quad  [\hat \phi_x,\hat \pi_y] = i  \frac{\delta_{xy}}{\sqrt{g_x}} \; .
\ee
The transverse lattice can
be restricted to finite volume with
a cut-off $\rho \le \rho_{\text{max}}$ as before.
Practically speaking there are worm cluster algorithms  appropriate for doing Monte
Carlo simulations for Ising and similar spins systems in continuous time \cite{Beard:1996wj}. 
This approach also provides the framework 
for going to Minkowski space
and the possibility of unitary algorithms
suited to a quantum computer.

\subsection{Continuous vs. discrete time}

To proceed with the Lagrangian  simulation we must 
discretize (\ref{eq:discreteAction}) with the
spacing 
$\Delta t = a_t$. Our Euclidean lattice action on AdS$_3$ is then
\begin{multline*} \label{eq:FullLatticeAction}
S = a_t  \sum_{x,t}  \Bigg[ \frac{1}{2} \sum_{y\in\<x,y\>}  \dfrac{\cosh \rho_x}{2} K_{xy}
(\phi_{x,t} - \phi_{y,t})^2 \\
+ \sqrt{g_x} \cosh \rho_x  \left( \frac{1}{2 a_t^2 \cosh^2 \rho_x} (\phi_{x,t}  - \phi_{x,t+1})^2 +  
\frac{1}{2} m^2 \phi^2_{x,t} + \lambda \phi^4_{x,t} \right) \Bigg] \; , \numberthis
\end{multline*}
with the lattice sites labelled by integer $x,t$. As in \cite{Brower:2019kyh}, we can make this expression more convenient by introducing the dimensionless parameters $m_0^2 = c_q^2 m^2$ and $\lambda_0 = 3 c_q^4 \lambda / q a_t A_{\Delta}$ in terms of an effective lattice spacing $c_q^2 = \sqrt{g_x} / K_{xy} = q a^2/4$. We are also free to choose the ratio of the spatial to temporal lattice spacing $a/a_t$. In this work we always set $a_t = c_q$ so that the coefficients of the spatial and temporal kinetic terms are the same. In the next section we will discuss the implications of this choice. After these substitutions and an appropriate rescaling of the field $\phi$, the lattice action becomes
\begin{multline*} \label{eq:SimpleLatticeAction}
S =   \sum_{x,t} \Bigg[\frac{1}{2} \sum_{y\in\<x,y\>}  \dfrac{\cosh \rho_x}{2}
(\phi_{x,t} - \phi_{y,t})^2 \\
+ \frac{1}{2 \cosh \rho_x} (\phi_{x,t}  - \phi_{x,t+1})^2 
+ \cosh \rho_x  \left(\frac{1}{2} m_0^2 \phi^2_{x,t} +  \lambda_0  \phi^4_{x,t} \right) \Bigg] \; . \numberthis
\end{multline*}

\begin{figure} 
\centering
\hspace{-1cm}
\includegraphics[width=\textwidth, angle=0, scale=0.6]{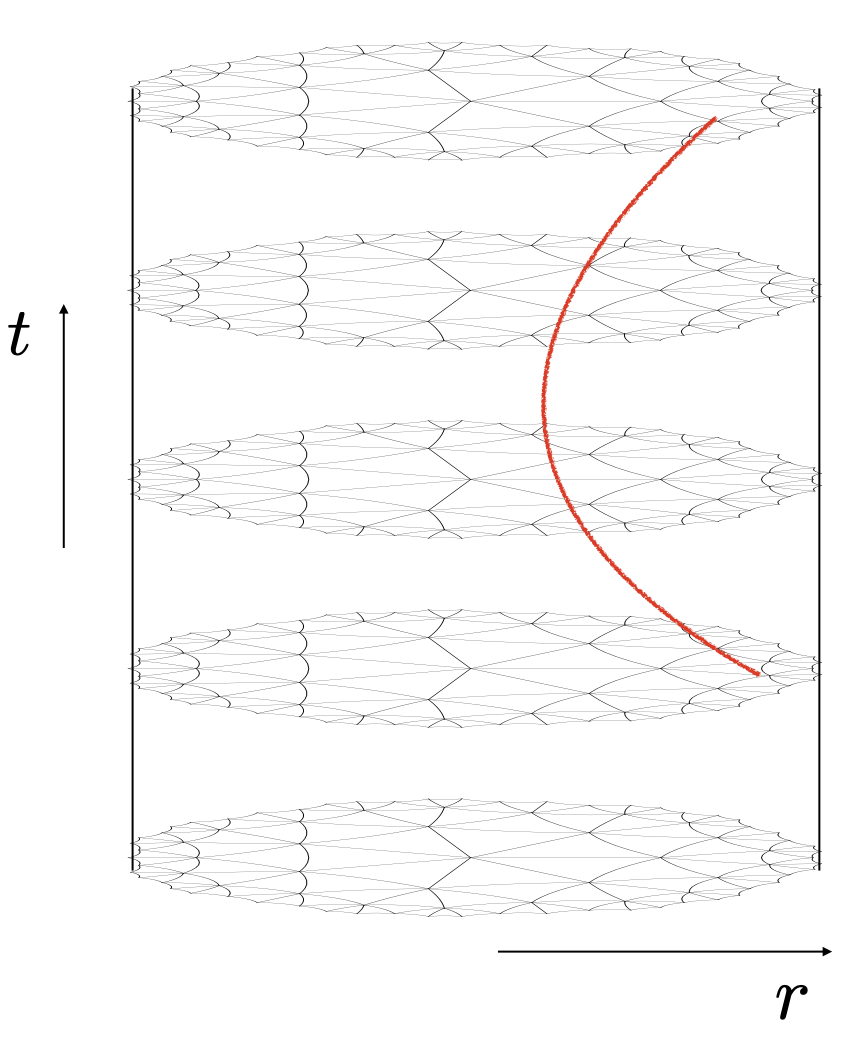}
\caption{A general bulk geodesic between two points bends into the bulk due to the time dilation of the coefficient $g_{00}(\rho)$
in the metric (\ref{eq:AdSDmetricBD}) when
moving toward the  boundary of  the AdS$_3$ cylinder. }
\label{fig:BulkGeo}
\end{figure}

It is important to note that because of the factors of $\cosh \rho_x$ in 
(\ref{eq:SimpleLatticeAction}),
the lattice weights (which were constant on the disk lattice) become position dependent on the AdS$_3$ lattice. Classically this comes from the $\cosh^2 \rho \, dt^2$ 
term in the metric (\ref{eq:globalAdS3metricHYP}) which indicates that there is a 
gravitational force pushing particles towards the center in this foliation due to the increased 
energy cost needed to move radially outwards (Fig.~\ref{fig:BulkGeo}).

We can check that the Hamiltonian (\ref{eq:discreteHam}) is consistent with  the
Lagrangian form (\ref{eq:FullLatticeAction}) in the limit that the temporal lattice spacing goes to zero. The   time-ordered partition 
function $Z = \Tr \exp[ - t \hat H]$  is then factorized into terms, 
\begin{align*}
Z_x = \< \phi_x(t+\delta t)|  e^{\textstyle{ - c_x  \hat \pi^2_x \delta t} / 2}  | \phi_x(t) \>
= \int d \pi_x \< \phi_x(t)|\pi_x\>  e^{\textstyle{ - c_x  \hat \pi^2_x \delta t / 2}}
  \<\pi_x | \phi_x(t + \delta t) \> \; , \numberthis
\end{align*} 
with
\be
c_x = \sqrt{g_x} \cosh \rho_x  \quad , \quad \< \phi_x(t)|\pi_x(t)\> = e^{ - i \sqrt{g_x}\pi_x \phi_x} \; .
\ee 
To understand the factor of $\sqrt{g_x}$ we rewrite the commutator using (\ref{eq:MomComm})
so that in flat space $\sqrt{g_x} \hat  \pi_x$ is the generator of translations in $\phi$. 
Completing the square gives
\be
Z_x =\int^\infty_{- \infty} d\pi_x \; e^{  i \pi_x \sqrt{g_x}[\phi_x(t+\delta t ) -\phi_x(t)] - c_x  \hat \pi^2_x \delta t / 2} 
\sim  \exp \bigg( - \frac{\delta t}{2}  \cosh \rho_x \sqrt{g_x} g^{00}_x  (\partial_t
\phi_x(t ))^2 \bigg) \; , \numberthis
\ee
as we would expect. 

\subsection{Lattice simulations}

The AdS$_3$ lattice is constructed by taking the hyperbolic disk 
lattice discussed in Sec.~\ref{sec:spatiallatice} with $L$ spatial layers and duplicating it to create $N_t$ time slices.
Dirichlet boundary conditions are imposed on a fictitious $(L+1)^{\text{th}}$ layer whereas periodic boundary 
conditions are taken in the time direction. 
Given that the AdS$_3$ lattice is an  extension of the hyperbolic disk lattice 
it shares many of the same properties.
Foremost, it shares the same exponential growth in points moving 
radially towards the boundary, as expected from holography:
$N_{tot} = N_t \times N_x \sim N_t \times e^{L}$, where 
$N_x$ is the number of spatial points. 
We do not refine the lattice for reasons similar to those discussed for the 2D case.

A crucial difference is that the lattice weights are now position-dependent, 
as discussed below (\ref{eq:SimpleLatticeAction}). A consequence of this is that when
traversing radially on the lattice towards the boundary, the time direction becomes exponentially stretched. This increases discretization effects close to the boundary, and makes probing the boundary theory a subtle task.
We can adjust this stretching by varying the ratio $a / a_t$,
which determines 
how stretched the temporal lattice spacing is relative to the spatial lattice spacing.
Because we are only studying bulk physics in this work, we fix this ratio for all simulations.
However, we note that in order to accurately explore the critical boundary theory it would be beneficial to vary this ratio to produce a more regular 
discretization of sites on the approach to the lattice boundary. We save this for future work. 

\section{The free theory} \label{sec:FreeTheory}
Since the free theory in the continuum has
a simple  analytical solution, we use it 
to check the fidelity of our lattice discretization
and the convergence of the  Monte Carlo simulation. 
For a given mass-squared $m^2$, the analytic bulk Green's function $G_{bb}(X, X')$ between two points 
$X$ and $X'$ in AdS$_{d+1}$ is the solution to the equation
\be \label{eq:GreenCont}
(-\nabla^2 + m^2)G = \frac{1}{\sqrt{g}} \delta^{d+1}(X-X') \; .
\ee
Here $\nabla_{\mu}$ is the covariant derivative and  $\nabla^2 = \nabla_{\mu} \nabla^{\mu}
= \frac{1}{\sqrt{g}} \partial_{\mu} \sqrt{g} g^{\mu \nu} \partial_{\nu}$ is the Laplace operator. 
The Green's function is given by \cite{Burgess:1984ti, Hijano:2015zsa}
\be \label{eq:FreePropCont}
G_{bb}(\sigma(X,X')) = e^{-\Delta \sigma} {}_2 F_{1} \bigg(\Delta, \frac{d}{2}, \Delta + 1-\frac{d}{2};
e^{-2 \sigma} \bigg) \; ,
\ee
where $\sigma$ is the geodesic between $X$ and $X'$ and the scaling dimension is related to the mass through 
$m^2 \ell^2 = \Delta(\Delta - d)$ with $\ell$ being the AdS radius. For $d=2$ the bulk Green's function in AdS$_3$ has the simple closed form
\be \label{eq:Gbb}
G_{bb}(\sigma) = \frac{e^{-\Delta \sigma}}{1-e^{-2\sigma}} \; ,
\ee
with the geodesic distance given by
\be
\cosh(\sigma) = \cosh(t - t') \cosh(\rho) \cosh(\rho') - \sinh(\rho) \sinh(\rho') \cos(\theta - \theta') \;
\label{eq:GeoDistance}
\ee
in global hyperbolic coordinates (\ref{eq:globalAdS3metricHYP}).
The free discretized  Green's function equation (\ref{eq:GreenCont})  satifies
the matrix equation  $M_{xt,\tilde{x} \tilde{t}} \, G(\tilde x ,\tilde t; x_0,t_0) = \delta_{x,x_0} \delta_{t,t_0}$ :
\bea
 \sum_{y\in \<x,y\>}  \frac{1}{2}(\sqrt{g^x_{00}} &+& \sqrt{g^y_{00}})(G(x,t;x_0,t_0)  -  G(y,t; x_0,t_0)) 
  +    m^2_0   \sqrt{g^x_{00}} G(x,t; x_0,t_0) \\
    &+&\frac{1}{\sqrt{g^x_{00}}}( 2  G(x,t;x_0,t_0) -  G(x,t+1;x_0,t_0) - G(x,t-1;x_0,t_0))   = \delta_{x,x_0} \delta_{t,t_0} \; , \nonumber
  \eea
  where $\sqrt{g^x_{00}}  = \cosh(\rho_x)$.
  This  is equivalent to the Gaussian  path integral $\<\phi(t,x)  \phi(t_0,x_0) \> \equiv  G(x, t;x_0,t_0)  $, allowing us to check the  Monte Carlo convergence  against the exact the matrix inverse $M^{-1}$ at $\lambda_0 = 0$.

For the massless case, we compute the lattice propagator $G_{\tilde{x} \tilde{t}, x_0 t_0} \equiv G(\tilde x ,\tilde t; x_0,t_0)$ via matrix inversion and compare it with the form of
the analytic propagator $G_{bb}$. We compute the lattice propagator betweean all lattice points, as well as
between only pairs of points with zero temporal or spatial separation. To avoid boundary effects, the $L^{\text{th}}$ layer is not included in measurements.
We check the Klebanov-Witten form 
$m^2 \ell^2 = \Delta(\Delta - d)$
from holography.
The results are shown in Fig.~\ref{fig:props} 
and show good agreement with the expected value of $\Delta = 2$ 
for the massless case.
We  note that similar to AdS$_2$ in \cite{Asaduzzaman_2020}, there is a small mass renormalization 
yielding an effective mass $m^2 > 0$ and a slightly larger scale, $\Delta > 2$.

\begin{figure} \label{fig:props}
\begin{center}
\includegraphics[width=0.49\textwidth]{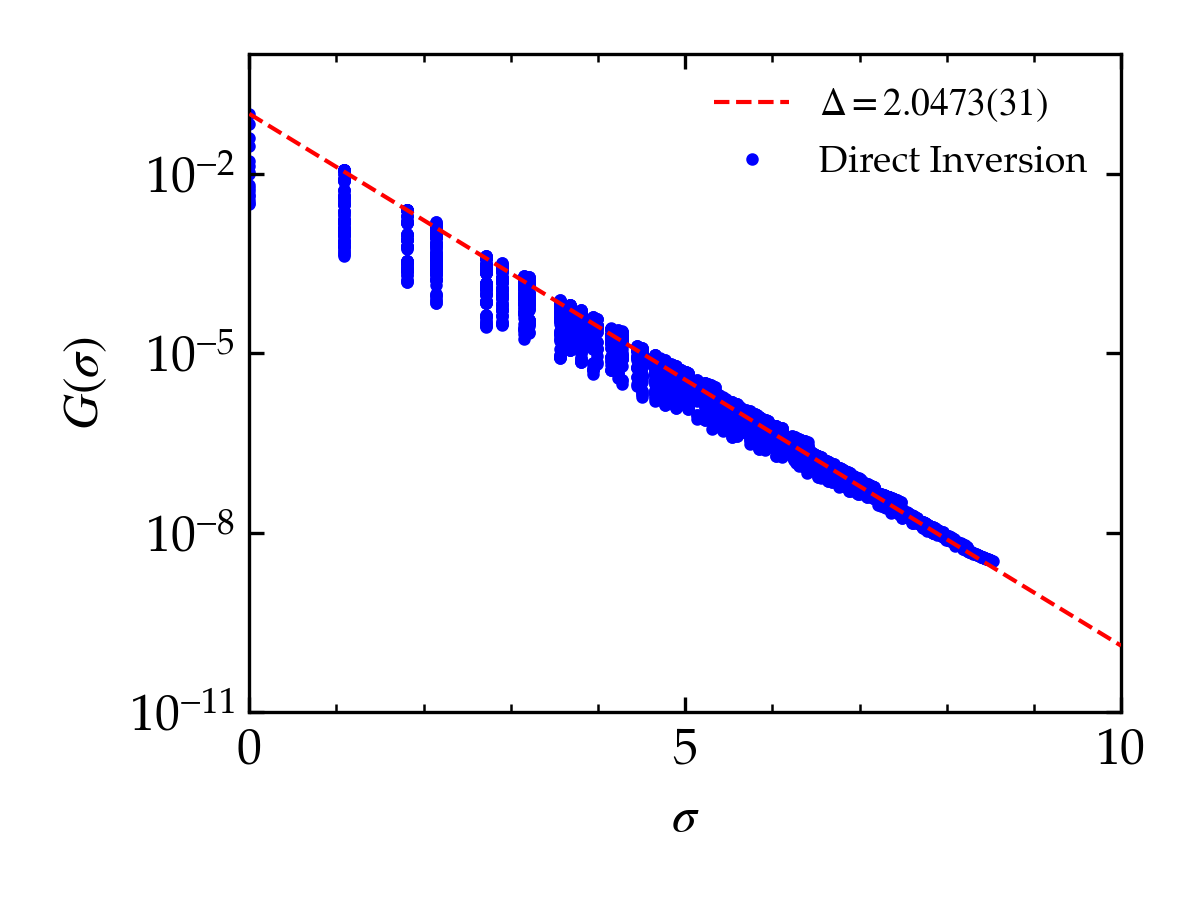}
\includegraphics[width=0.49\textwidth]{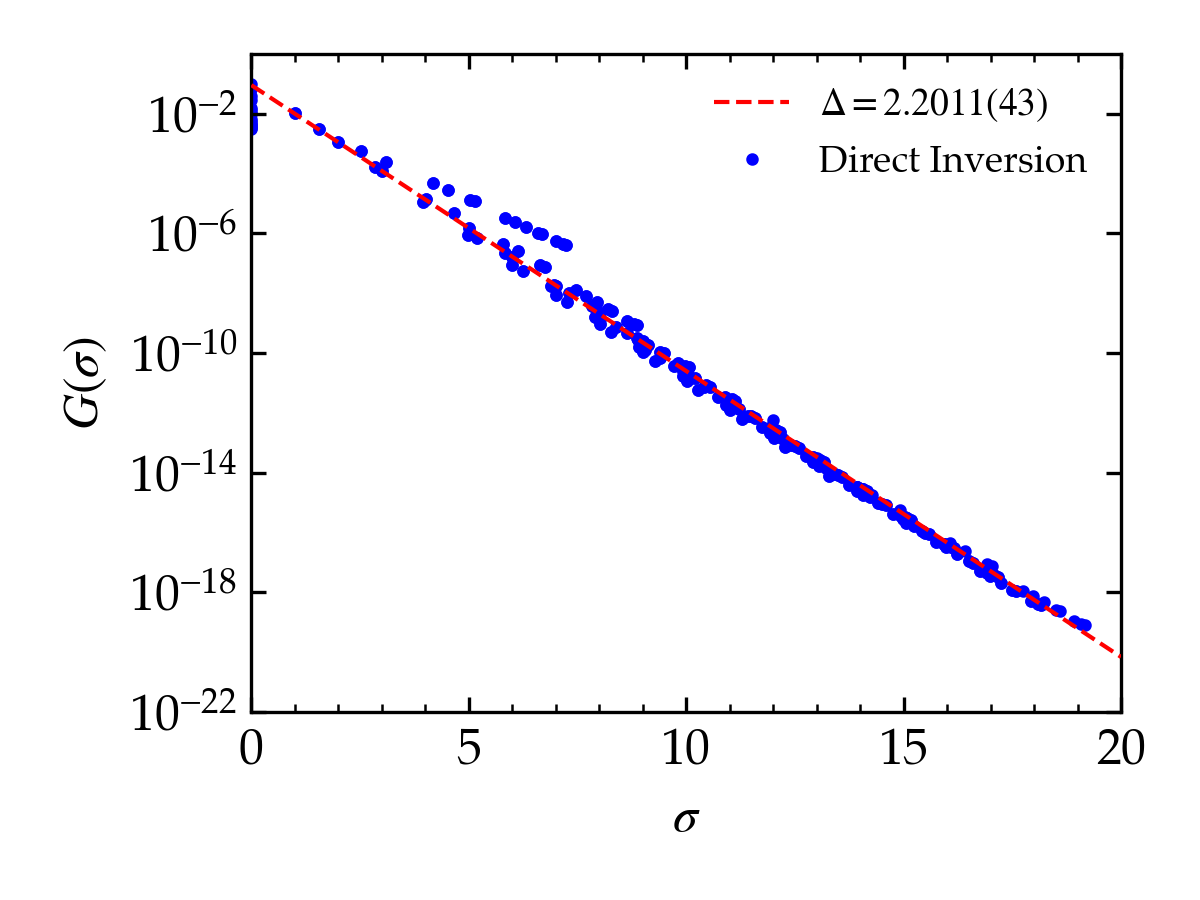}
\includegraphics[width=0.49\textwidth]{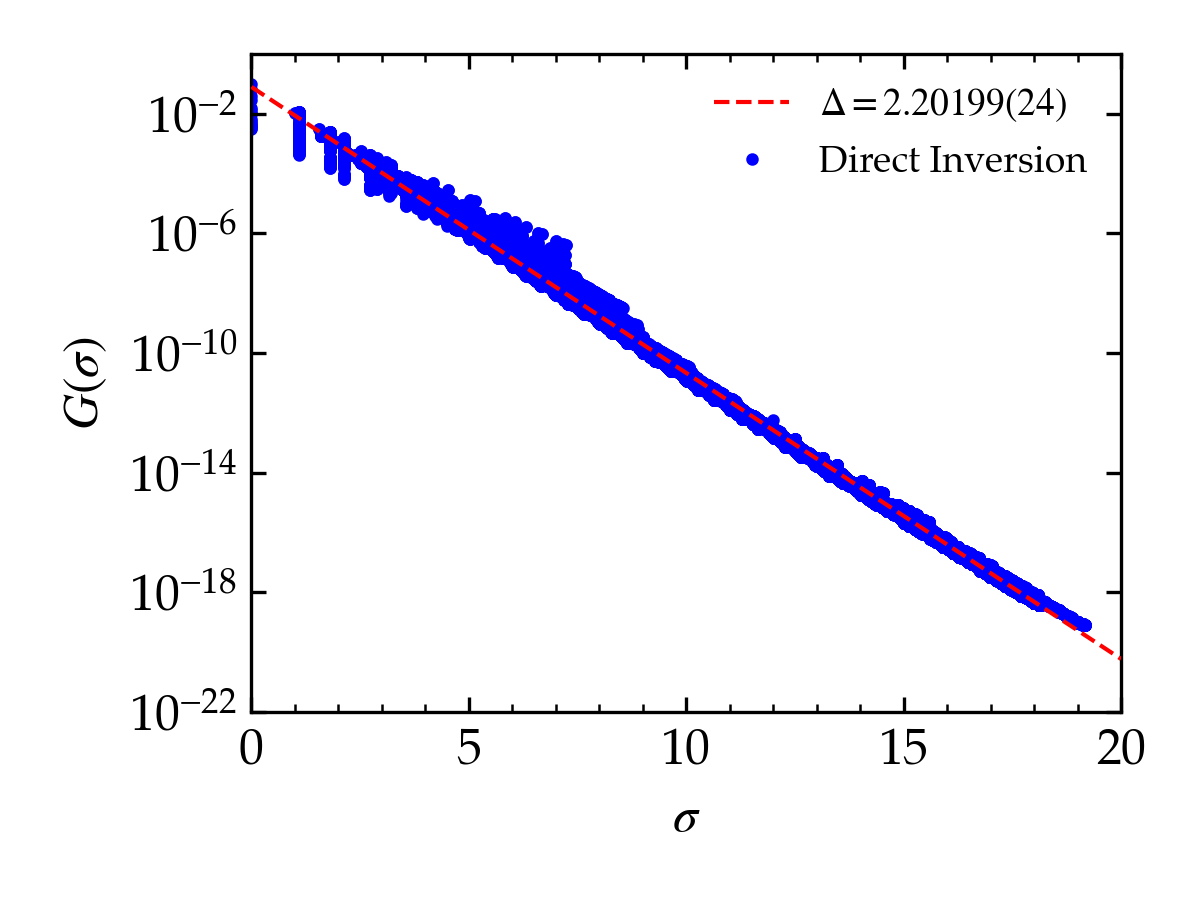}
\caption{Checks in the non-interacting regime of the AdS$_3$ lattice realization from a direct inversion 
of the massless Green's function $G_{bb}(\sigma)$ for $L = 4$. \textit{Top left:} The propagator for all pairs of points with zero temporal separation (i.e. pairs of points in the same time slice). \textit{Top right:} The propagator for all pairs of points with zero spatial separation. \textit{Bottom:} All-to-all propagator.}
\label{fig:props}
\end{center}
\end{figure}

In Fig.~\ref{fig:MCchecks} we compare the propagators from direct inversion to measurements from a Monte Carlo simulation of the same lattice. We see good agreement for all measurements larger than the statistical error in the Monte Carlo data, which is of order $10^{-6}$.

\begin{figure} \label{fig:MCchecks}
\begin{center}
\includegraphics[width=0.49\textwidth]{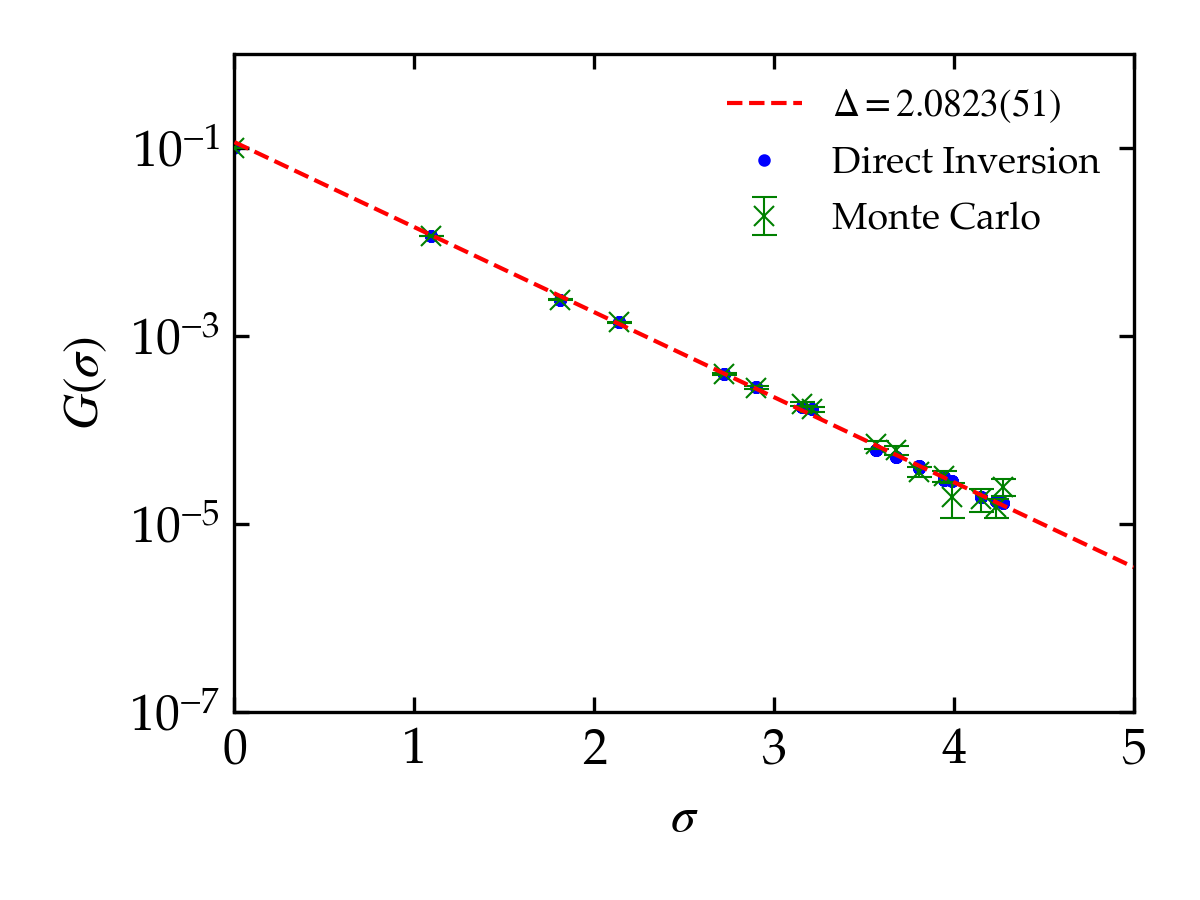}
\includegraphics[width=0.49\textwidth]{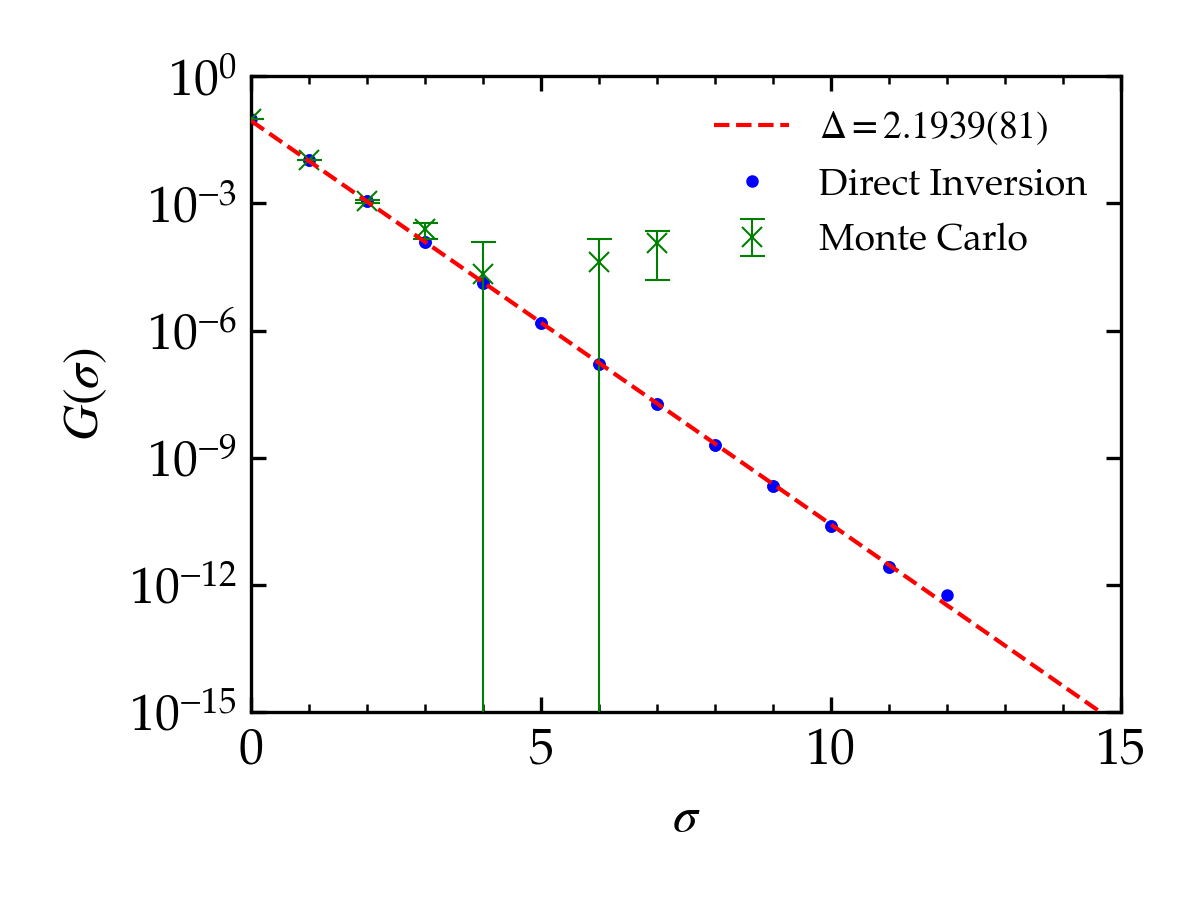}
\caption{Checks between the direct inversion and Monte Carlo for propagators 
for the massless case with $L = 4$. All of the fits shown are from the direct inversion data. \textit{Left:} Propagator from the center point to all other spatial points on the same time slice. \textit{Right:} Propagator along the center axis of the cylinder to all other temporal center points.}
\label{fig:MCchecks}
\end{center}
\end{figure}

\section{The interacting theory} \label{sec:MC}

To go beyond the free theory of Sec.~\ref{sec:FreeTheory} we use  the action (\ref{eq:SimpleLatticeAction}) 
with $\lambda_0 \neq 0$. 
We are specifically interested in determining if our lattice supports a critical point. This is the first step in being able to eventually determine what type of CFT is produced on the boundary at criticality.

\subsection{The $\phi^4$ critical point}
To study the theory (\ref{eq:SimpleLatticeAction}) with non-zero $\lambda_0$ and tachyonic mass $\mu_0^2 = -m_0^2$, 
we perform a lattice Monte Carlo simulation using a combination of Metropolis \cite{Metropolis1953Equation}, overrelaxation \cite{overrelax},
and the Brower-Tamayo cluster algorithm \cite{ Brower:1989mt}  with single cluster  Wolff  updates \cite{wolff}.
The critical point will depend on the two parameters, $\lambda_0$ and $\mu_0^2$, so we set $\lambda_0 = 1$ 
and sweep over $\mu_0^2$ values to find the critical $\mu^2_c$, which we define below. We repeat this process for an increasing number of lattice layers.
For each lattice size we choose the number of time slices $N_t$ to be equal to the number of  
points on the outermost spatial layer $L$ so that the lattice boundary has $N^2_t$ points. 

To analyze the critical behavior of the theory we measure two
bulk quantities: the magnetic susceptibility $\chi$ and
the Binder cumulant \cite{BINDER198879}. 
The magnetic susceptibility $\chi$ is defined as
\be
\chi = \dfrac{1}{V} \left( \<m^2\> - \<|m|\>^2 \right) \; ,
\ee
where we have introduced the lattice volume $V=N_t \sum_{x} \cosh \rho_x$ and the magnetization $m=\sum_{x,t} \cosh \rho_x \phi_{x,t}$. At a second-order phase transition we expect to see a peak in the susceptibility that grows with the lattice volume. The 
4th-order Binder cumulant 
$U_4$,
\be
U_4 = \dfrac{3}{2} \left( 1 - \dfrac{\langle m^4 \rangle}{3 \langle m^2 \rangle^2} \right) \; ,
\ee
 serves to determine whether the system is in the disordered phase $(U_4=0)$ or the ordered phase ($U_4=1$). At a second-order phase transition, we expect the Binder cumulant to approach a step function at the critical temperature as the lattice volume goes to infinity.

\begin{figure}[H]
\begin{center}
\includegraphics[width=0.49\textwidth]{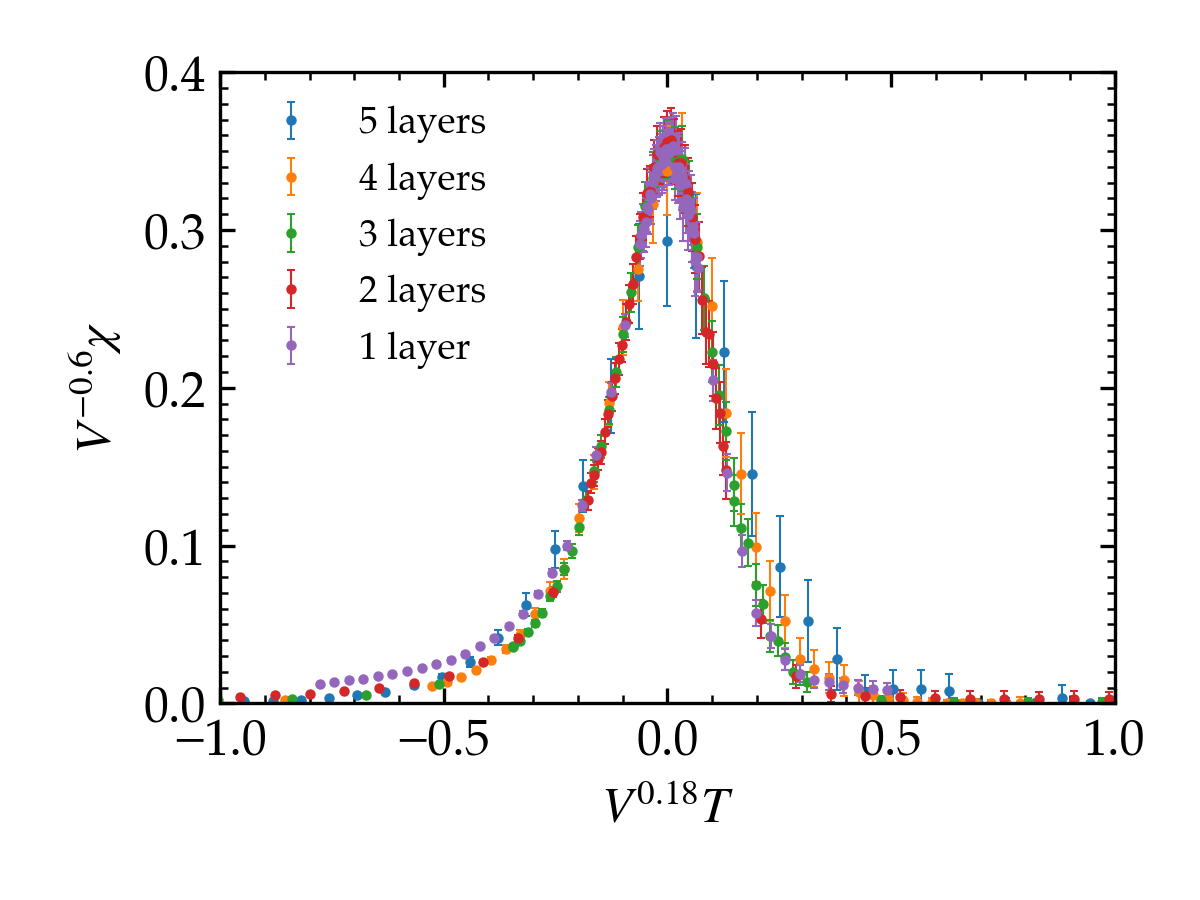}
\includegraphics[width=0.49\textwidth]{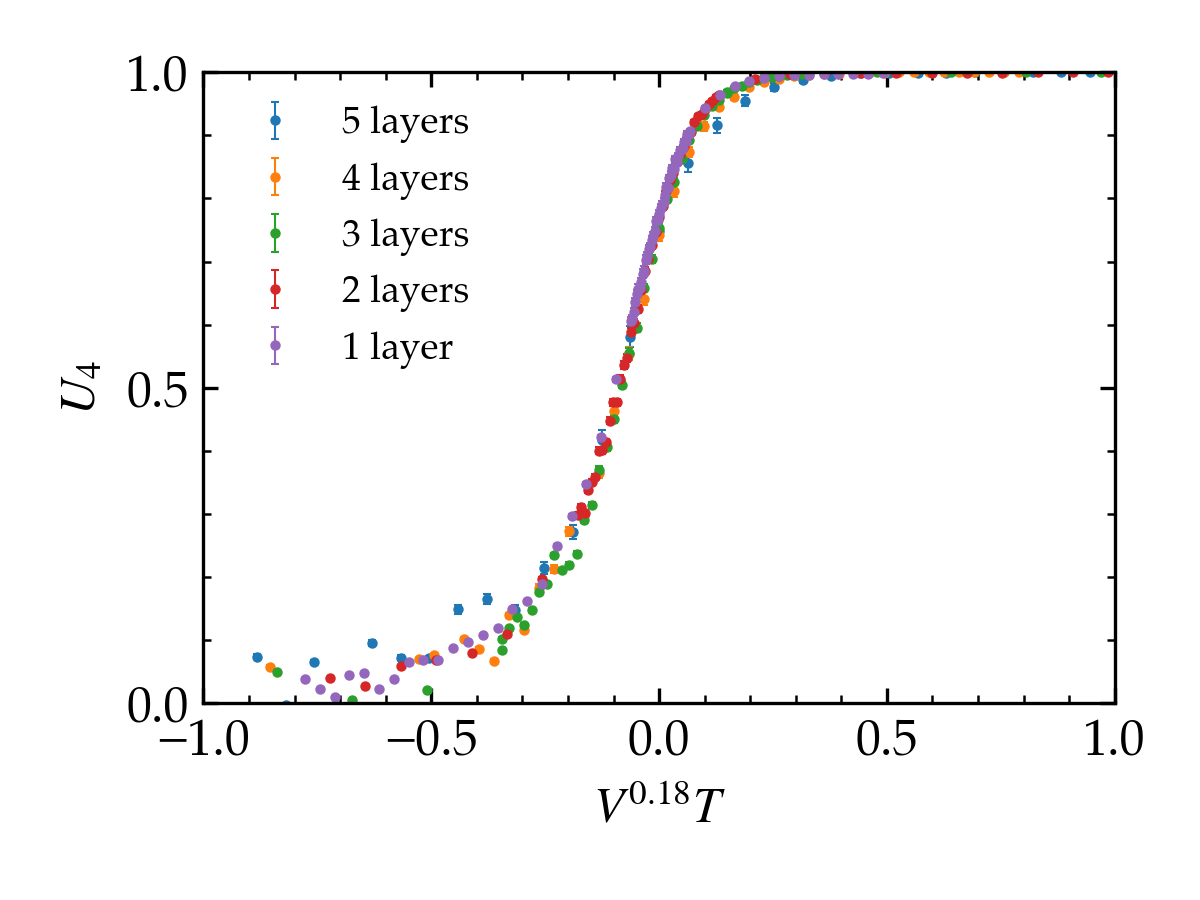}
\caption{Evidence for a second order phase transition for bulk $\phi^4$ theory in the 
Euclidean AdS$_3$ cylinder at $\lambda_0 = 1$.
\textit{Left:} The magnetic susceptibility $\chi$.
\textit{Right:} The 4th-order Binder cumulant $U_4$. We have scaled $\chi$ and $T$ by the lattice volume raised to an exponent, with the exponents chosen so that the data collapses onto a single curve.
}
\label{fig:CritPoint}
\end{center}
\end{figure}

We perform a finite-size scaling analysis \cite{fisher, landau, BINDER198879} by scaling $\chi$ and $T$ by powers of the volume, $V^{y_{\chi}}$ and $V^{y_{T}}$, respectively. We then adjust the exponents $y_{\chi}$ and $y_{T}$ so that the data for the different lattice sizes collapses onto a single curve, as shown in Fig.~\ref{fig:CritPoint}.
 The observed behavior is clear evidence of a second-order phase transition.
We note that the traditional finite-size scaling formalism is designed for uniform lattices in flat space, 
and without a careful discussion of finite size scaling in hyperbolic space we do not attempt to relate this scaling to conventional critical exponents for   either
a bulk transition or a transition for a  CFT at the boundary. Indeed,
since a large fraction of the total number of  lattice sites
are on the last layer at $\rho \simeq \rho_{\text{max}}$,  distinguishing between bulk and boundary criticality may be difficult at best; they 
may well be tied to each other as a consequence of
the AdS/CFT correspondence. We defer to a later work 
 the study of scaling for correlators and  
the goal to find the proper finite-size formalism
for AdS space.

We contrast the present discussion with the very interesting investigation in~\cite{Breuckmann_2020} on
the critical Ising model in hyperbolic space with periodic boundary
conditions.  With periodic boundary conditions,  these   hyperbolic triangulations
on closed Riemann manifolds in 2D are  the analogue of the Platonic solids on spheres 
for genus $g = 0$ and the finite triangulated torus for genus  $g =  1$.
For Riemann surfaces at higher genus ($g > 1$), the finite equilateral triangulations
correspond to negative curvature  manifolds.  The smallest classic example  is  the famous $g = 3$  Klein quartic \cite{Klein}  triangulated
by 56 $(2,3,7)$ equilateral triangles with  168 proper
symmetries.  Remarkably, this is the first in an infinite  sequence of larger volumes and
higher genuses \cite{Sausset_2007}.  

Given the highly connected graph  as the genus increases, it is  not surprising to find critical behavior with  mean field exponents as in \cite{Breuckmann_2020}.  This paper also finds
mean field exponents for the 3D  hyperbolic Ising lattice
with periodic boundary conditions. 
Moreover, periodic boundary conditions are interesting for applications to interacting particles in physical systems and necessary for experimentally realizing toric codes for quantum computing.  Nonetheless, this investigation~\cite{Breuckmann_2020} can not be directly compared with the critical properties seen in Fig.~\ref{fig:CritPoint}  on  our AdS tessellation approaching the hyperbolic surface  with  Dirichlet  boundary conditions at infinity. Instead this article  focuses  on the AdS/CFT correspondence with the goal to understand the non-perturbative relation between bulk and boundary critical behavior.

\section{Discussion} \label{sec:discussion}

Adapting the power of Euclidean lattice field theory to AdS space offers 
new possibilities for exploring strongly coupled phenomena. 
In this article we choose the simplest case
of scalar $\phi^4$ theory.   However, by implementing the
simplicial construction advocated in the Quantum Finite Element framework~\cite{brower2021PoS}, we believe
general field theories, including fermions \cite{Brower:2016vsl,Catterall:2018dns} and gauge fields \cite{Christ:1982ck} coupled to each other
or to scalars, can be realized on any smooth Riemann manifold.
By  constructing the lattice for AdS$_3$
 in a Euclidean cylinder geometry, we are able to  foliate  
 time with spatial sections on the Poincar\'{e} disk $\mH^2$ using
 the  triangle group \cite{Brower:2019kyh}, which is a discrete subgroup of the full conformal group.
The triangle group fixes the bulk UV lattice cut-off but the finite lattice volume  imposes
an arbitrary  center point  breaking the discrete isometries of AdS space.  However,
due the IR/UV connection of the AdS/CFT correspondence,
the IR cutoff $\rho_{\text{max}}$
implies an exponentially small UV cut-off $ \epsilon \sim \exp( - \rho_{\text{max}})$  for the  boundary CFT.
This raises the intriguing possibility of convergence to a continuum boundary CFT that does
not require UV completion in the bulk -- perhaps an echo of the basic concept of holography
for gauge/gravity duality.

We checked the free theory propagators using both direct matrix inversion as well as Monte Carlo methods and see 
scaling consistent with the continuum theory.  By looking at the magnetic susceptibility and the Binder cumulant we found strong evidence that there
is a bulk  critical point for  $\phi^4$ theory on our AdS$_3$ lattice.   Further simulations 
of the correlators 
are ongoing to  understand the subtleties of
the phase transition and to determine the approach to the CFT boundary (\ref{eq:CFTfield}).  
We seek to relate our finite volume scaling result to local
operators as well as address
the issue of the nature of the boundary CFT and whether it is a short- or long-range Ising model~\cite{Paulos:2016,Behan:2017emf}, or something else. To determine the scaling exponents requires more precision, but this is easily achieved. With efficient cluster algorithms \cite{ Brower:1989mt}, high statistics Monte Carlo simulations
for $\phi^4$ theory on lattice volumes up to $\mathcal{O}(10^6)$ sites
are feasible. Comparison with Table \ref{table:DiskInfo} implies that
lattices exceeding $L = 8$ with a number of time slices on the order of $N(L)$ are
reasonable. These
methods naturally apply to the Ising model,  which is presumably universally equivalent to $\phi^4$ theory approaching
the transition. 

Another feature of our lattice  is to introduce the Hamiltonian operator in Euclidean AdS space. For the
Ising representation we can treat time exactly using  continuous time loop algorithms \cite{Beard:1996wj}.
For both  AdS$_3$  and AdS$_2$ the Ising Hamiltonian takes the conventional form
\be
\hat H_{\mathrm{AdS}} = -\sum_i \cosh(\rho_i)\sigma^x_i -  t \sum_{\<ij\>} \cosh(\rho_i) \sigma^z_i \sigma^z_j \; ,
\ee
where the sum is over the transverse spatial links.  For AdS$_3$ the boundary  at $ \rho = \rho_{\max}$
is topologically a circle $\mS^1$ with $\theta \in [-\pi, \pi)$. For
AdS$_2$ this topology is even simpler: it is a strip~(\ref{eq:strip}), reminiscent 
of open strings but with 1D conformal quantum
mechanics at each end: $\theta =  0, \pi$. This geometry is amenable to many analytical methods
as highlighted in \cite{Hogervorst:2021spa}. It is also the 
simplest geometry in which to calculate bulk and boundary 
phenomena in the presence of differing boundary conditions 
\cite{Cogburn:2021awx}.
So both Hamiltonian AdS$_3$ and AdS$_2$ are ideal test systems.

The dynamics of the Ising model in hyperbolic space are more interesting and less well understood than in flat space, but
even the classical Ising model on $\mathbb H^2$ is interesting \cite{Krcmar:2008, Shima:2005vq, Asaduzzaman:2021bcw}.
Here the $K_{ij}$ weights for the regular triangle group lattice are constant, 
so in principle there is no need for counterterms to restore the
isometries at a second order phase transition. 
We note that
in (1+2)-dimensions we can use the Hamiltonian to go to Minkowski space and study unitary time
evolution $U(t) = \exp (-i t \hat H)$ suited to quantum computing.  One approach is
to simulate this on a digital quantum computer with the
standard Trotter expansion. 
An intriguing alternative is specific hardware being introduced 
\cite{Koll_r_2019, Bienias:2021lem, Boettcher:2019xwl, Yu:2020, Kollar:2019} that purports to realize
the discrete $\mathbb H^2$ lattice.

Finally,  time evolution for gravity is  an interesting
and challenging problem. Recently
 in Ref.~\cite{Cotler:2022weg}, similar finite element methods (FEM) 
were introduced in Minkowski space to address the
 interesting  problem of unitarity in an expanding universe. 
However, to enter the realm of gravity in our context requires dynamical metric
fluctuations for bulk gravity
dual to a boundary CFT with a conserved energy momentum
tensor. A natural framework is causal Regge calculus allowing for a change in
the simplicial geometry \cite{Catterall:2018dns}. With our construction a 
 first step is to consider weak gravitational fluctuations
around a fixed curved manifold by allowing the bonds
to fluctuate. 
This is yet another possible extension of this modest proposal in utilizing lattice field theory in an AdS background.

\section*{Acknowledgements}
We thank Liam Fitzpatrick, Ami Katz, and Chung-I Tan for helpful discussions.
This work was supported by the U.S. Department of Energy (DOE) under Award No.~DE-
SC0019139 and Award No.~DE-SC0015845.

\bibliographystyle{utphys}
\bibliography{ads3lat}{}

\end{document}